\documentclass[showpacs,amsmath,reprint,showkeys,pra,longbibliography]{revtex4-2}
\usepackage{dcolumn}    % Align table columns on decimal point
\usepackage{bm}         % bold math
\usepackage{hyperref}
\hypersetup{%
    colorlinks=true, 
    linkcolor=blue,
    urlcolor=blue,
    citecolor=blue, 
}
\usepackage{amssymb,lineno,amsfonts}
\usepackage{graphicx}   % Include figure files
\usepackage{bbm}
\usepackage{epstopdf}
\usepackage{setspace}
\usepackage{float}
\usepackage{natbib}
\usepackage{subfigure}
\usepackage{multirow}
\newcommand{\ket}[1]{\vert{#1}\rangle}
\newcommand{\bra}[1]{\langle{#1}\vert}

\begin{document}

 \title{Long-Lived Singlet State in an Oriented Phase and \\ its Survival across the Phase Transition Into an Isotropic Phase}

    \author{Vishal Varma}
    \email{vishal.varma@students.iiserpune.ac.in}
    \author{T S Mahesh}
    \email{mahesh.ts@iiserpune.ac.in}
    \affiliation{Department of Physics and NMR Research Center,\\
        Indian Institute of Science Education and Research, Pune 411008, India}

\begin{abstract}
Long-lived singlet states (LLS) of nuclear spin pairs have been extensively studied and utilized in the isotropic phase via liquid state NMR.  However, there are hardly any reports of LLS in the anisotropic phase that allows contribution from the dipolar coupling in addition to the scalar coupling, thereby opening many exciting possibilities. 
Here we report observing LLS in a pair of nuclear spins partially oriented in the nematic phase of a liquid crystal solvent.  The spins are strongly interacting via the residual dipole-dipole coupling.  We observe LLS in the oriented phase living up to three times longer than the usual spin-lattice relaxation time constant ($T_1$).
Upon heating, the system undergoes a phase transition from nematic into isotropic phase, wherein the LLS is up to five times longer lived than the corresponding $T_1$.  Interestingly, the LLS prepared in the oriented phase can survive the transition from the nematic to the isotropic phase. 
As an application of LLS in the oriented phase, we utilize its longer life to measure the small translational diffusion coefficient of solute molecules in the liquid crystal solvent. 
Finally, we propose utilizing the phase transition to lock or unlock access to LLS.
\end{abstract}

\maketitle
%~~~~~~~~~~~~~~~~~~~~~~~~~~~~~~~~~~~~~~~~~~~~~~~~~~~~~~~~~~~~~~~~~~~~~~~~~~~~~~~
%~~~~~~~~~~~~~~~~                                               ~~~~~~~~~~~~~~~~
%                           Section: Introduction
%~~~~~~~~~~~~~~~~                                               ~~~~~~~~~~~~~~~~
%~~~~~~~~~~~~~~~~~~~~~~~~~~~~~~~~~~~~~~~~~~~~~~~~~~~~~~~~~~~~~~~~~~~~~~~~~~~~~~~
\section{\label{sec:introduction} Introduction}
In nuclear magnetic resonance (NMR) spectroscopy, the longitudinal relaxation time constant ($T_1$) of nuclear spin states constrains the time 
during which a physical process can be studied. The discovery of long-lived singlet states (LLS), which have considerably longer lifetimes than $T_1$, has opened many novel applications \cite{2004_LLS_Levitt_1, book_pileio_2020}.
They include 
estimating slow diffusion rates of large biomolecules          \cite{2005_diffusion_Cavandini_Bodenhausen,2007_Sarkar_SS_exchange_NMR,2008_Sarkar_slow_diffusion_ASC,2009_Ahuja_diffusion_biomolecules}, 
storing hyperpolarized spin order \cite{2009_hyperpolarization}, 
detecting weak interactions between ligands and their binding sites for targeted drug deliveries 
\cite{2012_ligand_protein_screening,2014_ligand_protein_screening,2016_ligand_protein_screening}, 
studying slow chemical exchange  \cite{2021_chemical_exchange}, 
observing signals from hyperpolarized metabolites in magnetic resonance 
imaging \cite{2021_Metabolism}, and 
initialization of quantum registers for quantum information processing \cite{2010_SSRoy_Initialization_QRegisters}.
Various methods for LLS generation, efficient storage, precise manipulation, and robust detection have been described extensively in the literature \cite{Review_2019_LLS_Levitt,Review_2012_LLS_Levitt,book_pileio_2020}.

% Explain why singlet states have long lifetimes
% ---------------------------
The long lifetime of LLS arises from its immunity to intra-pair dipole-dipole relaxation, which is the major source of relaxation in NMR \cite{2005_LLS_Carravetta_Levitt,2010_Pileio_LLS_relaxation}. The transition from the antisymmetric singlet state to symmetric triplet states under the dipole-dipole interaction is symmetry forbidden. This gives LLS its extraordinarily long lifetime, even as long as an hour \cite{book_pileio_2020,2015_StevanatoG}. One generally exploits some form of asymmetry to populate and detect LLS. This asymmetry may be due to the chemical inequivalence of the spin-pair \cite{2004_LLS_Levitt_2,2011_LLS_Tayler_Levitt} or the magnetic inequivalence arising from differential coupling strengths with ancillary spins \cite{2012_LLS_Y.Feng,2014_YFeng_LLS}.

Various methods exist that generate LLS in different spin systems, such as the Carravetta-Levitt (CL) sequence for a weakly coupled spin-pair \cite{2004_LLS_Levitt_2}, M2S-S2M and SLIC sequence in the case of strongly coupled or chemically equivalent spin pair \cite{2010_LLS_Pileio_Levitt,2011_LLS_Tayler_Levitt,2012_LLS_Y.Feng,2013_SLIC_DeVience,2014_YFeng_LLS}, adiabatic methods \cite{2016_LLS_adiabatic}, and their variations \cite{2023_Abhinav_CD_LLS}. Discovering new methods for faster and more efficient generation of LLS \cite{2022_LLS_Aliphatic_Chains,2022_methylene_protons,2022_Selective_Excitation_LLS,2019_LLS_adiabatic,2018_LLS_optimal_control,2020_LLS_optimal_control} as well as synthesizing designer spin-systems that can sustain LLS for extraordinarily long durations \cite{2015_StevanatoG} are active areas of research.

In the isotropic phase (IP) of a liquid sample, the dipole-dipole interactions between two nuclear spins are completely averaged out due to the fast molecular tumbling, while the nonvanishing scalar coupling is generally small compared to the chemical shift difference.
Although it is easy to prepare LLS in such a weakly coupled spin system with strong chemical inequivalence, it needs to be sustained with the help of a symmetry-imposing spin-lock sequence \cite{2009_Pileio_spin_locking}. This limits
the storage time because of the heating caused by the spin-locking sequence. However, in a strongly coupled spin pair,  LLS closely resembles the system Hamiltonian's eigenstate and can be sustained without spin-lock  
\cite{2011_LLS_Tayler_Levitt,2012_LLS_Pileio,2013_LLS_Tayler_Levitt}.  

Although LLS has mostly been demonstrated in IP, we may also look for it in anisotropic phases.
The dipolar couplings of a spin-pair embedded in a crystalline lattice are too strong, which adds to the spectroscopic complexity \cite{book_levitt2013}. 
The partially oriented phase (POP) offers an excellent middle ground between the extreme cases of IP and crystalline phases. 
Nagashima et al \cite{nagashima2014long} have reported observing LLS in CH$_2$ protons of a tripeptide fused with a hydrogel.
Alternatively, the POP of a nematic liquid crystal solvent provides a convenient and controllable way to realize strongly coupled homonuclear spin-pairs. Liquid crystals have long been used as solvents in NMR spectroscopy for obtaining high-resolution spectra while partially retaining anisotropic interactions \cite{1963_Saupe_NMR_oriented, 1968_Meiboom_NMRinLC, 1968_Luckhurst_LC_solvent}. The residual dipole-dipole coupling of a spin system in POP can range from a few hundred Hz to a few kHz, depending on the order parameter of the liquid crystal matrix, which can be controlled via the sample temperature \cite{book_levitt2013}. 
Occurrences of POP in many biological systems, such as cell membranes, DNA, etc, have also been known for a long time \cite{2011_Jewell_BIOandLC, 1998_Goodby_BIOandLC, 1966_Stewart_BIOandLC}. Since NMR is a versatile tool for studying molecular transport in biological systems, the possibility of generating and sustaining LLS in such systems may significantly enhance the capability of NMR experiments.

\begin{figure}
\includegraphics[width=0.9\linewidth]{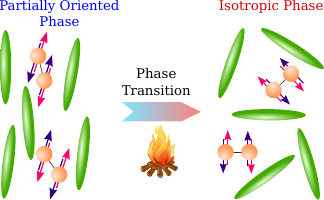}% 
\caption{\label{fig:abstract_image} LLS can not only be prepared in POP (left) but also can be carried across the phase-transition into IP (right).}
\end{figure}

In this work, we first show that the LLS of a two-spin system can be prepared and sustained in both the phases, POP as well as IP, of a nematic liquid crystal solvent.   
In NMR, the nuclear spin dynamics is mostly carried out in a particular 
thermodynamical phase of a sample. The finite lifetime of nuclear spin 
coherences is a hurdle for switching between thermodynamical phases of the 
physical system during a single transient. Our goal is to see if LLS prepared 
in POP can survive the phase transition into  IP brought about by heating the 
sample.  Interestingly, we find that LLS can survive the POP to IP phase 
transition of the liquid crystal solvent (see Fig. \ref{fig:abstract_image}).  
For this purpose, we first prepare LLS in POP, initiate the phase transition 
into IP by heating the sample, and finally convert LLS into observable 
magnetization in IP. We also utilize the long lifetime of LLS in these phases 
to estimate the translational diffusion coefficient at a set of temperatures. 

% how the article is organized?
% ---------------------------
This article is organized as follows. Sec. \ref{sec:theory} describes the theory, where we introduce the NMR Hamiltonian of a spin-pair in POP and explain the quantum state evolution during LLS generation, storage, and detection.
In the experimental section of Sec. \ref{ssec:lls_results}, we first discuss the experimental procedure to measure LLS lifetimes ($T_{LLS}$) at different temperatures and compare them with corresponding $T_1$.  Subsequently, we describe the results of experiments on the survival of LLS during POP to IP phase transition.  We then explain the experimental procedure and results for the estimation of diffusion coefficient via LLS in POP as well as IP. Finally, in Sec. \ref{sec:discussion_conclusion}, we discuss the significance of the experiments and make concluding remarks.

%~~~~~~~~~~~~~~~~~~~~~~~~~~~~~~~~~~~~~~~~~~~~~~~~~~~~~~~~~~~~~~~~~~~~~~~~~~~~~~~
%~~~~~~~~~~~~~~~~                                               ~~~~~~~~~~~~~~~~
%                           Section: Theory
%~~~~~~~~~~~~~~~~                                               ~~~~~~~~~~~~~~~~
%~~~~~~~~~~~~~~~~~~~~~~~~~~~~~~~~~~~~~~~~~~~~~~~~~~~~~~~~~~~~~~~~~~~~~~~~~~~~~~~
\section{\label{sec:theory}Theory}
\subsection{\label{ssec:pop_hamiltonian}The POP Hamiltonian}
Under a strong magnetic field $B_0$, the secular Hamiltonian in rotating frame for a nuclear spin-pair with scalar coupling $J$ as well as residual dipolar coupling $\mathcal{D}$ is given by (in $\hbar=1$ units)
\cite{book_levitt2013}
\begin{align}{\label{eq:th_Hamiltonian_general}}
    H &= -\pi \Omega I_{1z} + \pi \Omega I_{2z}
    + H_{12}, ~\mbox{where,}
    \nonumber \\
    H_{12} &= 2\pi J\mathbf{I}_1 \cdot \mathbf{I}_2
      + 2\pi \mathcal{D} \left(3I_{1z}I_{2z}  - \mathbf{I}_1\cdot\mathbf{I}_2
      \right).
\end{align}
Here $\Omega$ is the chemical shift difference between the two nuclear spins and $\mathbf{I}_i$ are the spin angular momentum operators with components $I_{i\alpha}$ with $\left(\alpha=x,y,z\right)$.  The residual  dipolar coupling $\mathcal{D}$ is the full dipolar coupling scaled by the order parameter of the POP, i.e., \(\mathcal{S} = \langle 3\cos^2(\Theta)-1\rangle/2\), wherein $\Theta$ is the angle between the inter-nuclear vector and the magnetic field $B_0$. The average $\langle~\rangle$ is taken over all possible orientations. In IP, the molecules exhibit random isotropic reorientations so that $\mathcal{S}$ and therefore $\mathcal{D}$ vanish.

\begin{figure}
\includegraphics[width=\linewidth]{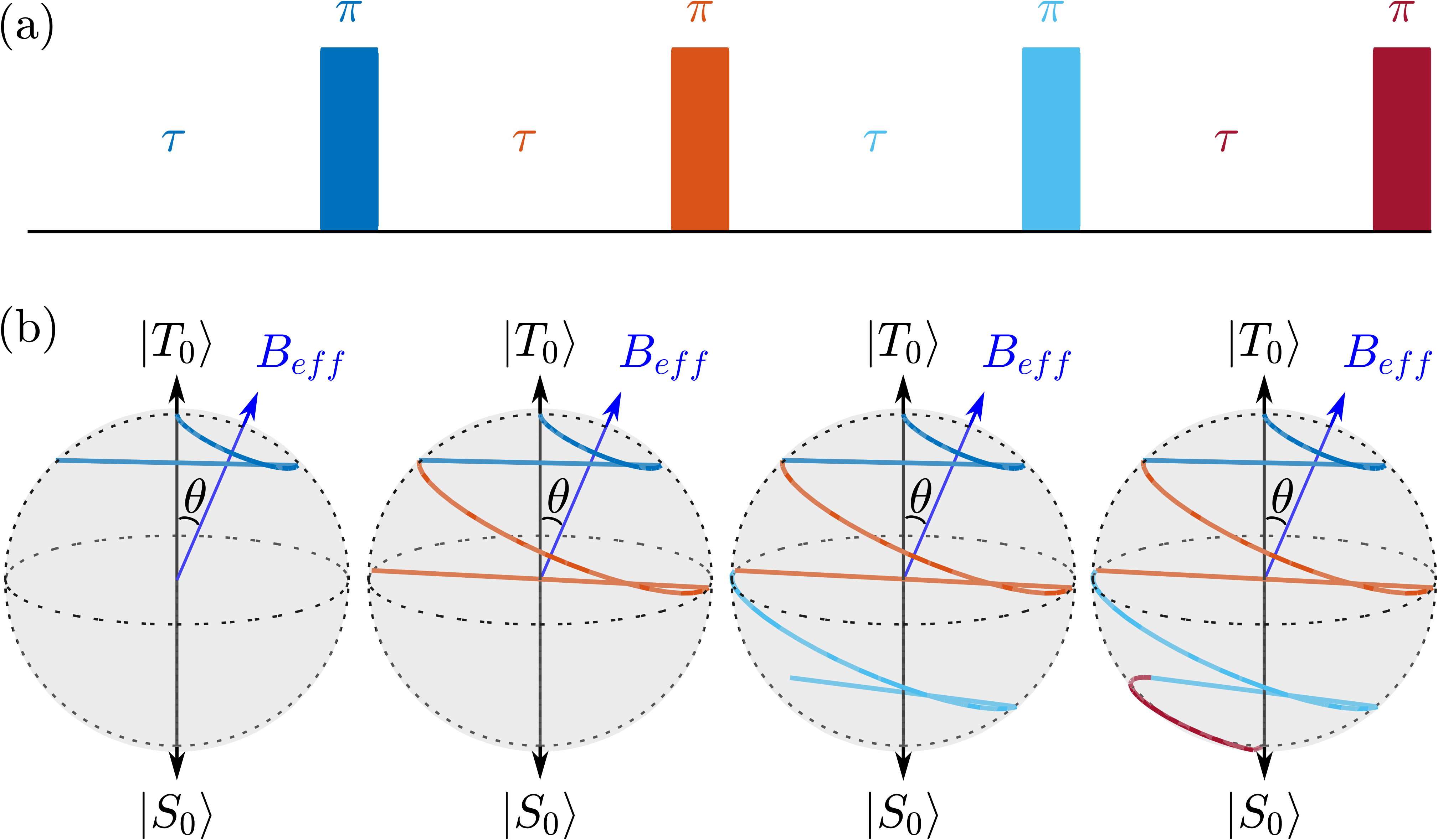}
\caption{\label{fig:ST_subspace_inversion} Rotations in $\{\ket{T_0},\ket{S_0}\}$ subspace by using CPMG echo sequence.
(a) CPMG Echo sequence. The delay $\tau$ and the number of iterations of CPMG echo are chosen to match the resonance condition (Eqs. \ref{eq:tau} and \ref{eq:n1}).
(b) Evolution from the initial state $\ket{T_0}$ under the resonant echo sequence.}
\end{figure}

\begin{figure*}
\includegraphics[width=0.95\linewidth]{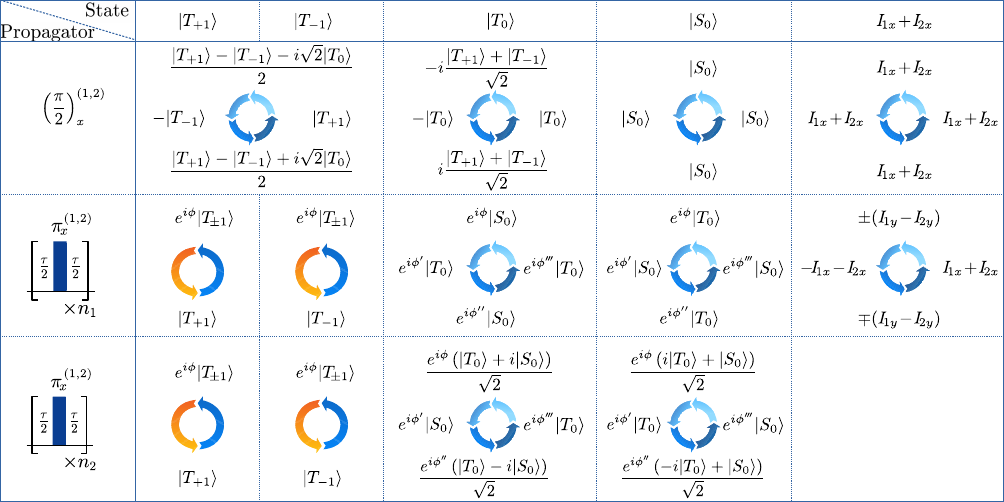}
\caption{\label{fig:ST_subspace_table} Evolution of singlet-triplet basis states under various elements of the M2S pulse sequence (see Fig. \ref{fig:m2s_s2m_pulses_and_evolution}). 
The column titles are the states that evolve under the operations listed in row titles.
Here $\tau$ given by Eq. \ref{eq:tau}.  The iteration numbers $n_1 = \lfloor \pi/(2\theta) \rceil$ and $n_2 = \lfloor \pi/(4\theta) \rceil$ respectively correspond to the resonant conditions for $\pi$ and $\pi/2$ rotations in $\{\ket{T_0},\ket{S_0}\}$ subspace.  The global phases are indicated by $\phi$, $\phi'$, $\phi''$, and $\phi'''$.
}
\end{figure*}

If the two spins are chemically equivalent (i.e., $\Omega=0$), either due to inherent symmetry or due to the suppression of the chemical shift by a spin-lock sequence such as WALTZ-16 \cite{book_cavanagh2007}, then
the Hamiltonian is simply $H_{12}$.
Consider the total spin angular momentum operator $\mathbf{S} = \mathbf{I}_1 + \mathbf{I}_2$ with quantum number $S$, and its z-component $S_z$ with the magnetic quantum number $m_S$. 
In the eigenbasis $\{\ket{0},\ket{1}\}$ of the Pauli $\sigma_z$ operator, the eigenstates of Hamiltonian $H_{12}$ are
\begin{align}
\ket{T_{+1}} &= \ket{S=1,m_S=1} = \ket{00}, 
    \nonumber \\
    \ket{T_0} &= \ket{S=1,m_S=0}=  \frac{1}{\sqrt{2}}(\ket{01} + \ket{10}),
    \nonumber \\
    \ket{T_{-1}} &= \ket{S=1,m_S=-1}= \ket{11},  
    \nonumber \\
    \ket{S_0} &= \ket{S=0,m_S=0}= \frac{1}{\sqrt{2}}(\ket{01} - \ket{10}).     
\end{align}
Singlet and triplet states have different exchange symmetries: $\ket{S_0}$ is 
anti-symmetric w.r.t. exchange, while $\ket{T_{m_S}}$ are symmetric.  
No symmetry-preserving operation can connect triplet and singlet states.  
For example, the intra-pair dipolar interaction commutes with the exchange operator and therefore can not induce transitions from $\ket{S_0}$ to  $\ket{T_{m_S}}$. Hence $\ket{S_0}$ gets  \textit{disconnected} from the rest of the triplet states and becomes the long-lived singlet state (LLS). 

In order to access LLS in terms of observable NMR magnetization, the intra-pair symmetry has to be broken. This breaking of symmetry is usually achieved by choosing a chemically inequivalent spin-pair ($\Omega\neq0$).
In a homonuclear spin-pair, the residual dipole-dipole coupling $\mathcal{D}$ is generally larger or comparable to the chemical shift difference $\Omega$.  The Hamiltonian $H$ for such a strongly coupled spin-pair can be conveniently expressed in the singlet-triplet basis as 
% \begin{equation}{\label{eq:H_in_ST_basis}}
% H = \dfrac{\pi}{2}
% \begin{bNiceMatrix}[first-row=0,last-col=5]
% \ket{T_{+1}}   & \ket{T_{0}}  & \ket{S_{0}}  & \ket{T_{-1}} & \\
% J+2\mathcal{D} & 0              & 0        & 0              & \bra{T_{+1}}\\\cline{2-3}
% 0              & J-4\mathcal{D} & -2\Omega & 0              & \bra{T_{0}}\\
% 0              & -2\Omega       & -3J      & 0              & \bra{S_{0}}\\\cline{2-3}
% 0              & 0              &  0       & J+2\mathcal{D} & \bra{T_{-1}}\\
% \end{bNiceMatrix}.
% \end{equation}
% \begin{equation}{\label{eq:H_in_ST_basis}}
% H = \dfrac{\pi}{2}
% \left[\begin{array}{cccc}
% J+2\mathcal{D} & 0              & 0        & 0             \\\cline{2-3}
% 0              & J-4\mathcal{D} & -2\Omega & 0             \\
% 0              & -2\Omega       & -3J      & 0             \\\cline{2-3}
% 0              & 0              &  0       & J+2\mathcal{D}
% \end{array}\right]
% \begin{array}{c}
% \bra{T_{+1}}\\ \bra{T_{0}} \\ \bra{S_{0}} \\ \bra{T_{-1}}\\
% \end{array}.
% \end{equation}
\begin{align}{\label{eq:H_in_ST_basis}}
H = \dfrac{\pi}{2}%
\begin{array}{c}
\begin{array}{cccccccc}
 \ket{T_{+1}} ~~~ & ~ \ket{T_{0}} ~~~ & \ket{S_{0}} ~~~ & 
\ket{T_{-1}} ~~~ & & & &
\end{array} 
\\
\left[\begin{array}{cccc}
J+2\mathcal{D} & 0              & 0        & 0             \\\cline{2-3}
0              & J-4\mathcal{D} & -2\Omega & 0             \\
0              & -2\Omega       & -3J      & 0             \\\cline{2-3}
0              & 0              &  0       & J+2\mathcal{D}
\end{array}\right]
\begin{array}{c}
\bra{T_{+1}}\\ \bra{T_{0}} \\ \bra{S_{0}} \\ \bra{T_{-1}}\\
\end{array}.
\end{array}
\end{align}
Two of the eigenstates of the Hamiltonian $H$ are simply the triplet states $\ket{T_{\pm 1}}$, whereas the other two eigenstates are linear combinations of $\ket{S_0}$ and $\ket{T_0}$ \cite{1957_Bernstein_spectra_analysis, book_levitt2013}.
Thus, the states $\{\ket{S_0},\ket{T_0}\}$ form an effective two-level subspace which can be conveniently represented by the Bloch sphere as shown in Fig. \ref{fig:ST_subspace_inversion}. The coupling term $-\pi\Omega$ can be used to interchange populations between $\ket{T_0}$ and $\ket{S_0}$ states.

%~~~~~~~~~~~~~~~~~~~~~~~~~~~~~~~~~~~~~~~~~~~~~~~~~~~~~~~~~~~~~~~~~~~~~~~~~~~~~~~
%~~~~~~~~~~~~~~~~~~~~~~~~~~~~~~~~~~~~~~~~~~~~~~~~~~~~~~~~~~~~~~~~~~~~~~~~~~~~~~~
\subsection{\label{ssec:theory_st_subspace}Rotations in Singlet-Triplet subspace}
The inversion of populations in \{$\ket{T_{0}}$, $\ket{S_{0}}$\} subspace is analogous to a single spin-1/2 system under RF drive \cite{2012_LLS_Y.Feng}.
In this analogy, (i) the difference of the two diagonal terms of the \{$\ket{T_{0}}$, $\ket{S_{0}}$\} subspace in Eq. (\ref{eq:H_in_ST_basis}) corresponds to an effective Larmor frequency $2\pi(J-\mathcal{D})$ 
and (ii) the sum of the off-diagonal terms corresponds to an effective RF field with amplitude $2\pi \Omega$ \cite{2011_LLS_Tayler_Levitt,2012_LLS_Y.Feng}. 
% Confusing label to (J-D)
The effective field in the \{$\ket{T_{0}}$, $\ket{S_{0}}$\} subspace makes an angle
\begin{align}
  \theta = \tan^{-1} \dfrac{\Omega}{J-\mathcal{D}}
\end{align}
with the $\hat{z}$ axis (see Fig. \ref{fig:ST_subspace_inversion} (b)).  Under this effective field of magnitude
$\nu_\mathrm{eff} = \sqrt{\Omega^2 + (J-\mathcal{D})^2},$
the state vector, initially pointed along $\ket{T_0}$, rotates around a cone by an angle of $2\theta$ in duration 
\begin{align}{\label{eq:tau}}
  \tau =  \dfrac{1}{2\nu_\mathrm{eff}}.
\end{align}
If a $\pi$ pulse is applied at this moment, the state flips to the other side of the Bloch sphere as indicated in Fig. \ref{fig:ST_subspace_inversion} (b). 
In the next $\tau$ duration, the state vector precesses along a new cone and reaches a maximum angle of $4\theta$ from the $\hat{z}$ axis.  
Continuing this way, a $\pi$ rotation in the \{$\ket{T_{0}}$, $\ket{S_{0}}$\} subspace can now be achieved by a resonant spin-echo transfer \cite{2011_LLS_Tayler_Levitt, 2012_LLS_Y.Feng}.  This inversion process is illustrated in Fig. \ref{fig:ST_subspace_inversion}.  
It involves repeated cycles of $\tau - \pi$ elements. % tau has not been defined yet.
Thus after $n$th iteration, the state vector is rotated by $2n\theta$.  The total number $n_1$ of iterations required to achieve inversion is given by
\begin{align}{\label{eq:n1}}
n_1 = \left\lfloor \dfrac{\pi}{2\theta} \right\rceil,
\end{align}
where $\lfloor \rceil$ denotes rounding to the nearest integer. In practice, one generally employs a CPMG sequence with repeated cycles of $(\tau/2 - \pi - \tau/2)$ elements, which also refocuses chemical shift.
Often $\pi$ pulses in the CPMG echo sequence are replaced by composite $\pi$ pulses that are robust against offset errors and RF inhomogeneity \cite{2010_LLS_Pileio_Levitt, 2011_LLS_Tayler_Levitt, 2012_LLS_Y.Feng}.
Some important basis state evolutions in the \{$\ket{T_{0}}$, $\ket{S_{0}}$\} subspace are illustrated in Fig. \ref{fig:ST_subspace_table}.

%~~~~~~~~~~~~~~~~~~~~~~~~~~~~~~~~~~~~~~~~~~~~~~~~~~~~~~~~~~~~~~~~~~~~~~~~~~~~~~~
%~~~~~~~~~~~~~~~~~~~~~~~~~~~~~~~~~~~~~~~~~~~~~~~~~~~~~~~~~~~~~~~~~~~~~~~~~~~~~~~
\subsection{\label{ssec:theory_m2s_s2m}Preparation, Storage, and Detection of LLS}
\begin{figure}
\includegraphics[width=\linewidth]{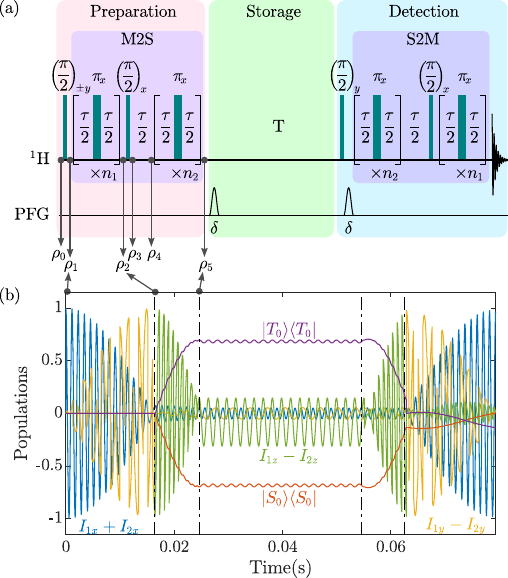}
\caption{ \label{fig:m2s_s2m_pulses_and_evolution}
(a) M2S-S2M pulse sequence for preparing, storing, and detecting LLS. 
Here $\tau$, $n_1$, and $n_2$ were calculated using Eqs. \ref{eq:tau} and \ref{eq:n1}. 
The two PFGs and the $\pi/2$ pulse in between are used to suppress artifacts.
After storing the LLS for duration T, it is converted to detectable $I_{1x}+I_{2x}$ magnetization using the S2M sequence. 
(b) Simulated evolution of the populations in states $\rho_1 = I_{1x}+I_{2x}$, $\rho_2 = I_{1y}-I_{2y}$, $\rho_3 =I_{1z}-I_{2z}$, $\ket{T_0}\bra{T_0}$, and $\ket{S_0}\bra{S_0}$ during the M2S-S2M pulse sequence assuming no relaxation and parameters $\Omega=50$ Hz, $J=10$ Hz, $\mathcal{D}=600$ Hz, T$=0.03$ s. 
Note that the LLS ($\rho_5$) persists throughout the storage interval without any spin-lock and is converted back to $\rho_1$ by the S2M sequence.
}
\end{figure}
The thermal equilibrium state of the homonuclear spin-pair of Larmor frequency $\omega_0$ is described by the Boltzmann distribution
\begin{align}
\rho_\mathrm{eq} 
\approx 
\dfrac{1}{2^n}\mathbbm{1} + \dfrac{\omega_0\beta}{2^n}\left(I_{1z} + I_{2z}\right)
\end{align}
where $\beta = 1/(k_B T)$. The identity part of the thermal density matrix can be ignored as this is invariant under any unitary operation \cite{book_levitt2013}. Therefore, the thermal density matrix in traceless deviation form is 
\begin{align}
\rho_0 = I_{1z} + I_{2z}.  
\end{align}

\begin{figure*}
\centering
\includegraphics{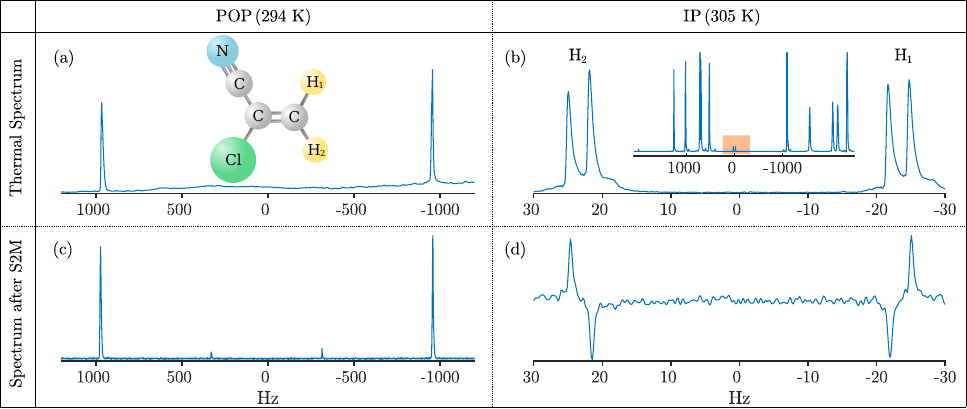}
\caption{\label{fig:molecule_and_simple_nmr_spectrum} (a,b) One-pulse $^1$H NMR spectrum of 2-Chloroacrylonitrile (CAN; molecular structure shown in the inset of (a)) in (a) POP at 294 K and (b) IP at 305 K (inset shows the full spectrum including the solvent peaks). (c,d) the spectrum after S2M 
in (c) POP at 294 K after a storage time of 7 s and (d) IP at 305 K after a storage time of 45 s.
}
\end{figure*}

The NMR pulse sequence for LLS preparation, storage, and detection is shown in Fig. \ref{fig:m2s_s2m_pulses_and_evolution} (a).
A $\left(\pi/2\right)_y$ pulse on $\rho_0$ creates single quantum coherence between the states $\ket{T_{\pm1}}$ and $\ket{T_0}$, i.e.,
\begin{align}
	\rho_1 &= I_{1x} + I_{2x} 
	= \dfrac{\ket{T_{+1}}+\ket{T_{-1}}}{\sqrt{2}} \bra{T_0}
 + \tt{h.c.},
\end{align}
where $\tt{h.c.}$ indicates the Hermitian conjugate term.
So far, the coherences are within the triplet subspace.
Now the CPMG echo train %$\frac{\tau}{2}-\pi-\frac{\tau}{2}$ 
$( \tau/2-\pi- \tau/2 )$ 
produces $\pi$ rotation in the \{$\ket{T_0}$, $\ket{S_0}$\} subspace as described earlier.  Using transformations in Fig. \ref{fig:ST_subspace_table} we obtain,
\begin{align}
  \rho_2 &= I_{1y} - I_{2y}
    = \dfrac{\ket{T_{+1}}+\ket{T_{-1}}}{\sqrt{2}} \bra{S_0} + \tt{h.c.}
\end{align}
The state $\rho_2$ has coherences of $\ket{S_0}$ with $\ket{T_\pm}$.  Now a $\left(\pi/2\right)_x$ pulse transfers  $(\ket{T_{+1}}+\ket{T_{-1}})/\sqrt{2}$ into $\ket{T_0}$ as shown in Fig. \ref{fig:ST_subspace_table}. This leads to a coherence between $\ket{S_0}$ and $\ket{T_0}$ states, i.e.,
\begin{align}
    \rho_3 = \ket{T_{0}}\bra{S_0} + \ket{S_{0}}\bra{T_0}  =
    I_{1z} - I_{2z}.
\end{align}
A further delay $\tau/2$ generates a relative phase shift in $\{\ket{T_0},\ket{S_0}\}$ subspace which is equivalent to a zero quantum coherence
\begin{align}
\rho_4 &=  i\left(\ket{T_0}\bra{S_0} - \ket{S_0}\bra{T_0}\right) =2I_{1y}I_{2x} -2I_{1x}I_{2y}.
\end{align}
Finally, a second CPMG echo train with $n_2 = \lfloor \pi/(4\theta) \rceil$ elements produces a $\left(\pi/2\right)$ rotation in the \{$\ket{T_{0}}$, $\ket{S_{0}}$\} subspace, converting the above coherence into population difference
\begin{align}
\rho_5 = \ket{S_{0}}\bra{S_0} - \ket{T_{0}}\bra{T_0},
\end{align}
which is the desired LLS.

A subsequent spoiling PFG (pulsed-field gradient), which is also a symmetry-preserving operation, can dephase unwanted coherences without affecting LLS.  As mentioned earlier, the LLS of a strongly coupled spin pair can be sustained in POP without any symmetry imposing spin-lock, which avoids unwanted sample heating.
The LLS detection is achieved by converting it back into observable magnetization using S2M, the reverse chronological ordered M2S sequence. Before S2M, a $\left(\pi/2\right)_y$ pulse followed by a spoiling PFG is  applied to remove any recovered longitudinal magnetization \cite{2012_LLS_Y.Feng}.

Fig. \ref{fig:m2s_s2m_pulses_and_evolution} (b) shows a numerical simulation of the evolution of different target states under the M2S--\textit{storage}--S2M pulse sequence. During the second CPMG echo train of M2S, the LLS (population difference between $\ket{T_0}$ and $\ket{S_0}$) starts to build up. Assuming no relaxation, the LLS persists during the storage interval and is converted back to observable magnetization by the subsequent S2M sequence.

%~~~~~~~~~~~~~~~~~~~~~~~~~~~~~~~~~~~~~~~~~~~~~~~~~~~~~~~~~~~~~~~~~~~~~~~~~~~~~~~
%~~~~~~~~~~~~~~~~                                               ~~~~~~~~~~~~~~~~
%                         Section: Experimets and Results
%~~~~~~~~~~~~~~~~                                               ~~~~~~~~~~~~~~~~
%~~~~~~~~~~~~~~~~~~~~~~~~~~~~~~~~~~~~~~~~~~~~~~~~~~~~~~~~~~~~~~~~~~~~~~~~~~~~~~~
\section{\label{sec:experiments}Experiments and Results}
In this work, our register involves two proton spins of 2-Chloroacrylonitrile (CAN) (see Fig. \ref{fig:molecule_and_simple_nmr_spectrum} (a)). The sample consists of a 137 mM solution of CAN in the nematic liquid crystal N-(4-Methoxybenzylidene)-4-butylaniline (MBBA). We observed the solution to be in the POP below 298 K and undergo a transition into the liquid phase at around 302 K. 
% density = 1.096 g/mL (25 deg Celcius)
% mol. weight = 87.51 g/mol
% solute concentration : ~ 8 uL
% solvent concentration : ~ 730 uL
% moles of solute (n) : (solute_vol*density)/mol_weight = 0.1002e-3 mol
% Molarity = 0.1002e-3/730e-6 = 0.1373 M
All experiments were performed on a 500 MHz Bruker AVANCE-III NMR spectrometer operating with a static magnetic field of strength 11.7 Tesla.

Fig. \ref{fig:molecule_and_simple_nmr_spectrum} (a) shows the one-pulse NMR spectrum of CAN at 294 K, while it is in the POP, in which the two large peaks correspond to the two outer peaks of a strongly dipolar-coupled spin pair \cite{book_levitt2013}. The two middle peaks are undetectably small and are lost in the liquid crystal background signal.  
Fig. \ref{fig:molecule_and_simple_nmr_spectrum} (b) shows the one-pulse spectrum at 305 K when the solution is in IP. From this spectrum, we estimated the chemical shift difference $\Omega=46.6$ Hz and indirect scalar coupling $J=3.1$ Hz.

\begin{figure}
\centering
 \includegraphics{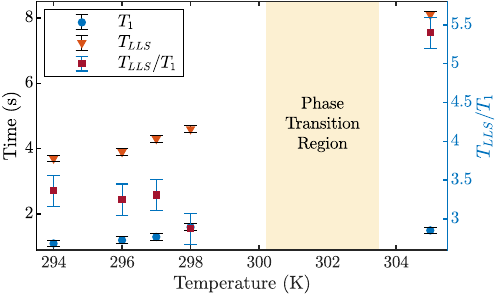}
\caption{\label{fig:lls_results} 
Measured values of $T_1$,  $T_{LLS}$, and their ratio $T_{LLS}/T_1$ for CAN at different temperatures across both phases. $T_{LLS}$ was obtained by systematically varying the storage delay T in Fig. \ref{fig:m2s_s2m_pulses_and_evolution} (a), while $T_1$ was measured using the standard inversion-recovery method \cite{book_cavanagh2007}. In POP, $T_{LLS}$ is approximately three times longer than $T_1$, whereas it is five times longer in IP.}
\end{figure}

\subsection{\label{ssec:lls_results} Establishing LLS in POP and IP}
We used the M2S-S2M \cite{2010_LLS_Pileio_Levitt,2012_LLS_Y.Feng} pulse sequence shown in Fig. \ref{fig:m2s_s2m_pulses_and_evolution} to prepare, store, and detect LLS in POP. 
The resulting spectrum, shown in Fig. \ref{fig:molecule_and_simple_nmr_spectrum} (c), consists of four almost equally spaced lines with characteristic intensities corresponding to the singlet state, from which the estimated residual dipole-dipole coupling constant is $\mathcal{D} \approxeq 640$ Hz.
% new edit
The value of $\mathcal{D}$ was estimated by using the four signal intensities ($I_1$, $I_2$, $I_3$, $I_4$) and the transition frequencies ($\nu_1$, $\nu_2$, $\nu_3$, $\nu_4$) from the spectrum in Fig. \ref{fig:molecule_and_simple_nmr_spectrum}c, and then running an optimization routine to find the best value that satisfied the equations 
\begin{align}
\nu_1 = \nu_{o\!f\!f} + ( J + 2\mathcal{D} + C)/2; &~~ I_1 = (1 - \sin\xi)/2 \nonumber \\
\nu_2 = \nu_{o\!f\!f} + ( J + 2\mathcal{D} - C)/2; &~~ I_2 = (1 + \sin\xi)/2 \nonumber \\
\nu_3 = \nu_{o\!f\!f} + (-J - 2\mathcal{D} + C)/2; &~~ I_3 = (1 + \sin\xi)/2 \nonumber \\
\nu_4 = \nu_{o\!f\!f} + (-J - 2\mathcal{D} - C)/2; &~~ I_4 = (1 - \sin\xi)/2,
\end{align}
where
\( C = \sqrt{(J-2\mathcal{D})^2 + \Omega^2} \), 
\( \tan \xi = (J-\mathcal{D})/\Omega \), and $\nu_{o\!f\!f}$ is the RF offset frequency \cite{book_levitt2013}.
The spin pair is strongly coupled in POP, but as we raise the sample temperature, $\mathcal{D}$ decreases, and finally, as the solution undergoes a phase transition into IP at around 302 K, $\mathcal{D}$ averages out completely, and spins become weakly coupled under the scalar coupling $J$ with $|J| \ll |\Omega|$.  
To prepare and read LLS in the IP, we used the CL-CL pulse sequence given in \cite{2004_LLS_Levitt_2}, which results in a characteristic antiphase signal as shown in Fig. \ref{fig:molecule_and_simple_nmr_spectrum} (d). 
Since the singlet is no longer an eigenstate of the weakly coupled Hamiltonian, we used a 1 kHz WALTZ-16 spin lock during the storage interval. 
The measured $T_1$, $T_{LLS}$, and the ratio $T_{LLS}/T_1$ at various temperatures spanning both POP and IP are plotted in Fig. \ref{fig:lls_results} and summarized in Table \ref{table:all_results}.
The results clearly establish the long-livedness of the singlet state not only in IP but also in POP.

\begin{center}
  \begin{table}
    \caption{Summary of experimental results for CAN at various temperatures spanning POP and IP.}
    \centering	
    \begin{tabular}{|c|c|c|c|c|c|}
      \hline
       & Temp. & $T_1$(s) & $T_{LLS}$(s) & \multicolumn{2}{c|}{Diffusion Coefficient}\\
      Phase & (K) &  & & \multicolumn{2}{c|}{$D$($\times10^{-10}m^2s^{-1}$)}\\\cline{5-6}
      & &          &            & STE & LLS\\\hline
      %--------------------------------------------------------------------%
      %T  &  T1         &     Tlls    &       Dse     &      Dlls    \\\hline
      %--------------------------------------------------------------------%
      \multirow{4}{*}{POP}& 294 & 1.1$\pm$0.1 & 3.7$\pm$0.1 & 1.29$\pm$0.17 & 1.32$\pm$0.10 \\
      \cline{2-6}
      %--------------------------------------------------------------------%
      & 296 & 1.2$\pm$0.1 & 3.9$\pm$0.1 & 1.34$\pm$0.12 & 1.34$\pm$0.14\\ \cline{2-6}
      %--------------------------------------------------------------------%
      & 297 & 1.3$\pm$0.1 & 4.3$\pm$0.1 & 1.37$\pm$0.13 & 1.37$\pm$0.11\\ \cline{2-6}
      %--------------------------------------------------------------------%
      & 298 & 1.6$\pm$0.1 & 4.6$\pm$0.1 & 1.55$\pm$0.21 & 1.45$\pm$0.12\\ \hline
      %--------------------------------------------------------------------%
      IP & 305 & 1.5$\pm$0.1 & 8.1$\pm$0.1 & 1.81$\pm$0.03 & 1.92$\pm$0.13\\
      \hline
    \end{tabular}
	\label{table:all_results}
  \end{table}
\end{center}

%~~~~~~~~~~~~~~~~~~~~~~~~~~~~~~~~~~~~~~~~~~~~~~~~~~~~~~~~~~~~~~~~~~~~~~~~~~~~~~~
%                         Sub-section: Phase Transition
%~~~~~~~~~~~~~~~~~~~~~~~~~~~~~~~~~~~~~~~~~~~~~~~~~~~~~~~~~~~~~~~~~~~~~~~~~~~~~~~
\subsection{\label{ssec:phase_transition} Survival of LLS across the phase transition}
\begin{figure}
\centering
\includegraphics{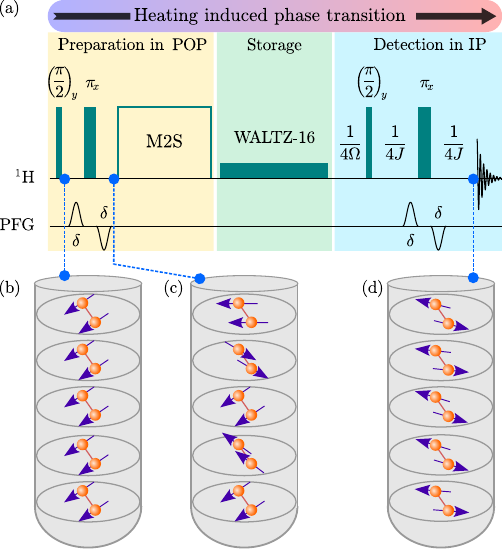}
\caption{\label{fig:lls_across_phase_transition}
(a) Pulse sequence for preparing LLS in POP using M2S sequence shown in Fig. \ref{fig:m2s_s2m_pulses_and_evolution} and detecting it after the phase transition into IP using modified CL sequence. Before LLS preparation in POP, the first bipolar-PFG introduces a z-dependent phase shift, which is refocused by the second bipolar-PFG after the phase transition into IP. This way, the `stimulated echo' signal is solely due to the LLS that survived the phase transition. 
In all experiments, sinusoidal PFG of duration $\delta=320\mu$ s were used with strength 2.5 G/cm. Since the system is transitioning from strong coupling to weak coupling, we use a 2 kHz WALTZ-16 spin-lock to sustain LLS during storage.
(b-d) Illustrating spin states (b) after the first $(\pi/2)_y$ pulse when both spins are pointing along $\hat{x}$, (c) after the first bipolar-PFG, and (d) $I_{1y}-I_{2y}$ state after the phase transition into IP. 
}
\end{figure}
As explained in Sec. \ref{ssec:theory_m2s_s2m},  M2S  \cite{2010_LLS_Pileio_Levitt}  efficiently transfers the longitudinal magnetization to LLS in a strongly coupled spin pair in POP, whereas the CL sequence \cite{2004_LLS_Levitt_2} is efficient for converting LLS back to the observable magnetization of the weakly coupled spin pair in IP. Thus in this experiment, we introduce a hybrid M2S-CL sequence as shown in Fig. \ref{fig:lls_across_phase_transition} (a).

In addition, we introduce two important improvements as described below. 
Firstly, to remove spurious contributions to the final signal, we need an efficient phase-cycling scheme. 
Since the phase transition is not rapidly reversible, it has to be a single-scan phase cycle. To this end, we incorporate stimulated-echo sequence with the help of two PFG pairs, one during preparation before the M2S sequence and the other during detection after the CL sequence. 
Together, they filter-in only the signal that arises from the LLS and suppress all artifacts created during storage or at other times. 
We call this sequence  \textit{M2S-CL STELLAR (STimulated Echo filtered Long-Lived state Accessing and Reading)}. A vectorial illustration of this sequence is shown in Fig. \ref{fig:lls_across_phase_transition} (b).
Secondly, the rapid nonuniform heating and the associated phase transition render the solution highly inhomogeneous, resulting in the broadening of the spectral lines. Consequently, the characteristic anti-phase spectral lines of Fig. \ref{fig:molecule_and_simple_nmr_spectrum} (d) vanish under line broadening. Therefore, we introduce a spin-echo sequence in the detection part to convert the anti-phase magnetization $I_{1x}I_{2z}-I_{1z}I_{2x}$ into in-phase magnetization $I_{1y}-I_{2y}$. 

\begin{figure}
\centering
\includegraphics[width=\linewidth]{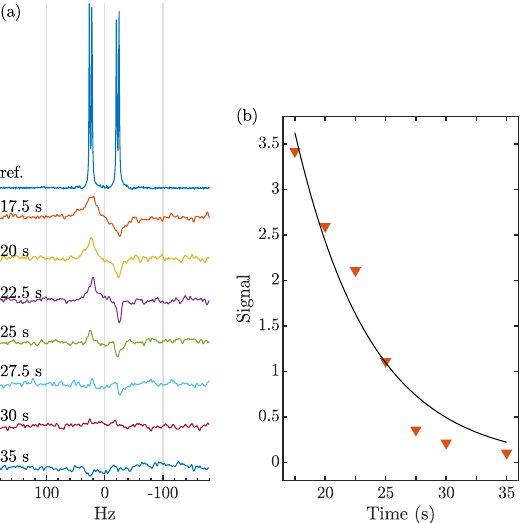}
\caption{\label{fig:phase_transition_results} 
(a) Spectra corresponding to LLS that has survived the phase transition at various storage intervals as mentioned. Here the top trace is reference, same as in Fig. \ref{fig:molecule_and_simple_nmr_spectrum} (b).
(b) Decay of LLS signal intensities in (a) versus storage interval $T$.
}
\end{figure}

Fig. \ref{fig:phase_transition_results} (a) shows the spectra corresponding to $I_{1y}-I_{2y}$ magnetization obtained from LLS that has survived the phase transition  occurred during storage intervals of different durations. 
Fig. \ref{fig:phase_transition_results} (b) plots the decay of the survived LLS signal versus storage time. In this trans-phase storage, we obtained an effective LLS lifetime of 6.3 s, which lies in between the values obtained for individual phases.

% new edit
The efficiency of M2S-CL STELLAR sequence was numerically estimated to be about 40\% (under instantaneous phase transition from POP to IP and with no storage delay). In our experiments, we could retain roughly 2\% of the magnetization with 17.5 s of storage time, which amounts to the pulse-sequence efficiency of about 30\%.

%~~~~~~~~~~~~~~~~~~~~~~~~~~~~~~~~~~~~~~~~~~~~~~~~~~~~~~~~~~~~~~~~~~~~~~~~~~~~~~~
%                    Sub-section: Diffusion experiments
%~~~~~~~~~~~~~~~~~~~~~~~~~~~~~~~~~~~~~~~~~~~~~~~~~~~~~~~~~~~~~~~~~~~~~~~~~~~~~~~
\subsection{\label{ssec:diffusion_results} Estimating diffusion coefficient in POP and IP}
DOSY (diffusion-ordered spectroscopy) is an established NMR technique to study molecular diffusion from small molecules to large polymers \cite{2017_diffusion_review}.  Its principle can be explained by considering an ensemble of molecules, each with a single spin-1/2 nucleus. 
After preparing $(\ket{0}+\ket{1})/\sqrt{2}$ state by an initial $\pi/2$ pulse, a PFG introduces a local phase shift to prepare $(\ket{0}+e^{i\phi(z)}\ket{1})/\sqrt{2}$. A reverse PFG is applied after a sufficiently long diffusion interval. In the absence of diffusion, the local phase shift is completely reversed, and one obtains a strong echo signal. In the presence of translational diffusion, the phase reversal is inefficient and the echo signal is damped. 
For a fixed diffusion interval $\Delta$, PFG strength $G$, and duration $\delta$, the signal ratio is given by \cite{Review_2001_diffusion}
\begin{equation}\label{eq:diffusion_ratio_general}
S(G)/S(0) = \exp\left(-D \kappa^2 \Delta\right).
\end{equation}
Here $D$ is the diffusion coefficient, $\kappa = \gamma q G \delta s$ with $\gamma$ being the gyromagnetic ratio, $q$ being the coherence order, and $s$ being PFG shape-factor \cite{book_cavanagh2007}.
Thus, $D$ can be estimated by  measuring $S(G)/S(0)$ for varying $G$ and fitting with
the Gaussian function above. 
In practice, the precision of this method is limited by the hardware bound on $G$ and the coherence-time bound on $\Delta$. 
\begin{figure}
\centering
\includegraphics[width=\linewidth]{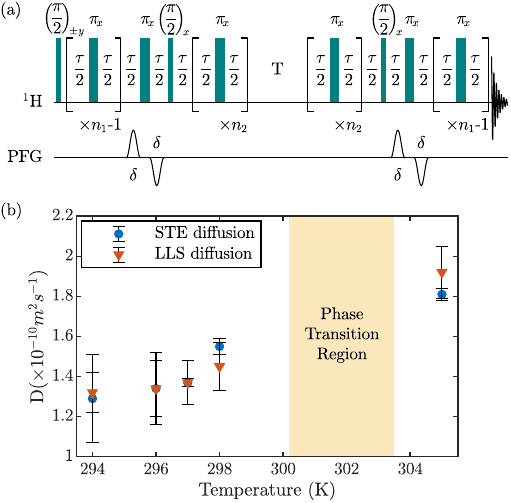}
\caption{\label{fig:diffusion_pulse_and_results} (a) Pulse sequence used for estimation of diffusion coefficient using LLS. 
The parameters $\tau$, $n_1$ and $n_2$ are optimized at each temperature for maximum final signal. 
Position encoding and decoding are realized by bipolar-PFGs (of duration $\delta = 320 \mu$ s) to minimize eddy currents. %\cite{2005_diffusion_Cavandini_Bodenhausen}.
For STE diffusion experiment at each temperature, PFG strength $G$ was varied from 2.5 G/cm to 47.5 G/cm in 19 steps,  with diffusion interval $\Delta=3.3$ s. 
For LLS diffusion experiment in POP, $\Delta = 10$ s and $G$ was same as in STE;  whereas in IP, $\Delta = 30$ s and $G$ was varied from 1 G/cm to 20 G/cm 20 steps.
(b) The measured translational diffusion coefficient of CAN in MBBA plotted versus temperature using STE sequence \cite{2005_diffusion_Cavandini_Bodenhausen} and 
PGM2S2M sequence shown in (a). 
}
\end{figure}
Cavadini et al \cite{2005_diffusion_Cavandini_Bodenhausen} proposed the LLS method for studying slow diffusion, which was later applied also to strongly coupled systems \cite{2017_pileio_pgM2S, 2018_sad_nmr}.
Indeed, for slow diffusion studies with strongly coupled spin-pairs, such as those in POP, LLS is ideal because it can be sustained over long diffusion intervals without spin-lock. 

Here we measure $D$ for CAN in MBBA at a range of temperatures spanning both POP and IP using LLS as well as the conventional stimulated echo (STE) methods. For POP, we use the pulse sequence shown in Fig. \ref{fig:diffusion_pulse_and_results} (a), which was referred to as PGM2S2M \cite{2017_pileio_pgM2S}. 
 For IP, we use the CL-CL based DOSY sequence by Cavadini et al \cite{2005_diffusion_Cavandini_Bodenhausen}.  For comparison, we also measured the diffusion coefficient using the conventional STE method at all temperatures.
The results shown in Fig. \ref{fig:diffusion_pulse_and_results} (b) and also summarized in Tab. \ref{table:all_results} indicate a gradual increase in the diffusion coefficient with temperature, as expected.

%~~~~~~~~~~~~~~~~~~~~~~~~~~~~~~~~~~~~~~~~~~~~~~~~~~~~~~~~~~~~~~~~~~~~~~~~~~~~~~~
%~~~~~~~~~~~~~~~~                                               ~~~~~~~~~~~~~~~~
%                      Section: Discussion and Results
%~~~~~~~~~~~~~~~~                                               ~~~~~~~~~~~~~~~~
%~~~~~~~~~~~~~~~~~~~~~~~~~~~~~~~~~~~~~~~~~~~~~~~~~~~~~~~~~~~~~~~~~~~~~~~~~~~~~~~
\section{\label{sec:discussion_conclusion} Discussions and Conclusions}
\begin{figure}
\centering
\includegraphics[width=\linewidth]{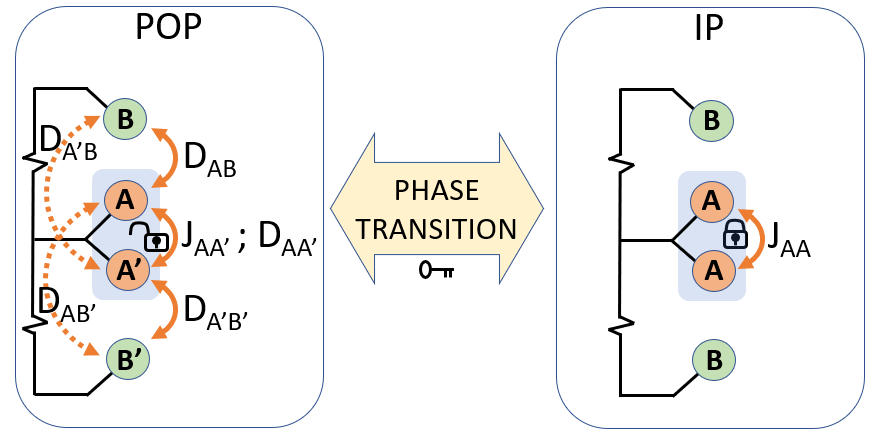}
\caption{\label{fig:lls_unlock_pt} Proposed PADLOCK method using a pair of A spins and covalently farther B spins as shown.  In POP, due to the heteronuclear dipolar coupling, the magnetic equivalence is broken, and LLS in the AA'BB' system can be accessed.  In IP, the dipolar coupling vanishes leaving only scalar coupling between magnetically equivalent AA spins, which potentially can hold LLS for indefinitely long times. Thus, the phase transition is the key to accessing or locking LLS.
}
\end{figure}
Since their discovery two decades ago, long-lived singlet states (LLS) have opened a plethora of applications, from precision spectroscopy to medical imaging.  However, LLS have been mostly observed in isotropic phases, wherein the anisotropic interactions such as dipolar couplings are averaged out.  Here we reported the observation of LLS in a spin-pair of a solute in the partially oriented phase (POP) of a liquid crystal solvent.
To observe LLS in such a strongly dipolar coupled system, we used the M2S-S2M pulse sequence, originally designed for strongly $J$-coupled systems in isotropic phase (IP).  We analyzed the related spin dynamics by constructing rotation elements of the singlet-triplet basis states as well as their populations under each element of the pulse sequence.
In the particular spin pair that we studied, the LLS lifetime in POP was observed to be three times longer than the longitudinal relaxation time constant $T_1$. Heating the solution takes it to IP, rendering the spin pair weakly coupled. Using the relevant CL-CL sequence, we observed LLS and measured its lifetime, which was found to be about five times longer than $T_1$.

The observation of LLS in POP naturally raises an interesting question of whether LLS survives a phase transition of the solution from POP to IP.  
To investigate this point, we introduced the M2S-CL STELLAR sequence, a hybrid of M2S and CL, with single-scan LLS filtering via stimulated echo technique. 
Using this sequence, we have been able to prepare LLS in POP, store it during the phase transition, and sensitively detect it in IP. The experimental results  not only revealed the survival of LLS across the phase transition but also yielded an effective LLS time-constant which happens to be about four times the average $T_1$ value.  
Finally, we demonstrated an application of LLS by efficiently measuring the translational diffusion coefficient at various temperatures spanning both phases. 

% application in NMR-QIP
The presence of large yet manageable residual dipolar couplings without severely compromising coherence times makes spin systems in POP attractive for a variety of spectroscopic as well as quantum information processing applications.  Many biological systems, such as membrane proteins, are also found in POP in their cellular conditions \cite{2022_review_bio_LiquidCrystal}. Therefore, the observation of LLS in POP may extend the breakthroughs of precision spectroscopy beyond IP.
Strong dipolar couplings in POP are helpful not only for preparing LLS via polarization transfer or algorithmic cooling \cite{2017_AlgoCool_Deepak,2020_AlgoCool_LLS_Levitt}, but also for sustaining LLS without external fields.
We may also envisage other novel applications such as investigating dynamics during phase transition as well as hybridizing the merits of two phases for spectroscopy. For example, we may ask if LLS prepared in POP be safely locked in IP. This possibility, which we refer to as PADLOCK (Phase transition Assisted Detection and LOCKing), is illustrated in Fig. \ref{fig:lls_unlock_pt}.

\section{Acknowledgements}
We acknowledge valuable discussions with V. R. Krithika, Priya Batra, and Dr. Sandeep Mishra. 
The funding from DST/ICPS/QuST/2019/Q67 is gratefully acknowledged. 
We also thank the National Mission on Interdisciplinary Cyber Physical Systems for funding from the DST, Government of India through the I-HUB Quantum Technology Foundation, IISER-Pune.  

% \nocite{*}
\bibliography{references}

\end{document}